\documentclass[journal=jacsat,10pt,manuscript=article,layout=twocolumn]{achemso}

\usepackage{mathtools, cuted}
\usepackage{caption}

\usepackage{lineno,hyperref}
\modulolinenumbers[5]
\usepackage{amssymb}
\usepackage{times}
\usepackage{graphicx}
\usepackage[usenames,dvipsnames,svgnames,table]{xcolor}
\usepackage{natbib}
\usepackage{array}
\usepackage{tabularx}
\onecolumn

\author{J. Palot\'{a}s}
\affiliation{Department of Physics, Budapest University of Technology and Economics and
MTA-BME Lend\"{u}let Spintronics Research Group (PROSPIN), POBox 91, H-1521 Budapest, Hungary}
\alsoaffiliation{Radboud University, Institute for Molecules and Materials, FELIX Laboratory, Toernooiveld 7, 6525ED Nijmegen, The Netherlands}

\author{M. Negyedi}
\affiliation{Department of Physics, Budapest University of Technology and Economics and
MTA-BME Lend\"{u}let Spintronics Research Group (PROSPIN), POBox 91, H-1521 Budapest, Hungary}
\alsoaffiliation{Universit\"{a}t T\"{u}bingen Physikalische Institut, Auf der Morgenstelle 14D 72076 T\"{u}bingen, Germany}
\alsoaffiliation{HighFinesse GmbH, Auf der Morgenstelle 14D 72076 T\"{u}bingen, Germany}

\author{S. Kollarics}
\affiliation{Department of Physics, Budapest University of Technology and Economics and
MTA-BME Lend\"{u}let Spintronics Research Group (PROSPIN), POBox 91, H-1521 Budapest, Hungary}

\author{A. Bojtor}
\affiliation{Department of Physics, Budapest University of Technology and Economics and
MTA-BME Lend\"{u}let Spintronics Research Group (PROSPIN), POBox 91, H-1521 Budapest, Hungary}

\author{P. Rohringer}
\affiliation{University of Vienna, Faculty of Physics, Strudlhofgasse 4., Vienna, A-1090, Austria}

\author{T. Pichler}
\affiliation{University of Vienna, Faculty of Physics, Strudlhofgasse 4., Vienna, A-1090, Austria}

\author{F. Simon}
\email{f.simon@eik.bme.hu}
\affiliation{Department of Physics, Budapest University of Technology and Economics and
MTA-BME Lend\"{u}let Spintronics Research Group (PROSPIN), POBox 91, H-1521 Budapest, Hungary}
\alsoaffiliation{Laboratory of Physics of Complex Matter, \'{E}cole Polytechnique F\'{e}d\'{e}rale de Lausanne, Lausanne CH-1015, Switzerland}

	\title{Incidence of Quantum Confinement on Dark Triplet Excitons in Carbon Nanotubes}

\begin{document}
	
\begin{abstract}
Photophysics of single-wall carbon nanotubes (SWCNTs) is intensively studied due to their potential application in light harvesting and optoelectronics. Excited states of SWCNTs form strongly bound electron-hole pairs, excitons, of which only singlet excitons participate in application relevant optical transitions. Long-living spin-triplet states hinder applications but they emerge as candidates for quantum information storage. Therefore knowledge of the triplet exciton energy structure, in particular in a SWCNT chirality dependent manner, is greatly desired. We report the observation of light emission from triplet state recombination, \textit{i.e.} phosphorescence, for several SWCNT chiralities using a {\color{black}purpose-built} spectrometer. This yields the singlet-triplet gap as a function of SWCNT diameter and it follows predictions based on quantum confinement effects. Saturation under high microwave power (up to 10 W) irradiation allows to determine the spin-relaxation time for triplet states. Our study sensitively discriminates whether the lowest optically active state is populated from an excited state on the same nanotube or through F\"{o}rster exciton energy transfer from a neighboring nanotube. 
\end{abstract}

\section*{KEYWORDS:}{carbon nanotubes, optically detected magnetic resonance, relaxation times, quantum confinement, molecular rulers, F\"{o}rster exciton transfer}

\maketitle

\maketitle
\clearpage
\newpage
\twocolumn

Understanding the photophysics of novel materials is crucial for their effective application in both light harvesting and light generation, which are relevant for diverse fields including energy applications, optoelectronics, telecommunications, photonics, and quantum technology. The fundamental processes involve either the generation or the recombination of an electron-hole pair. This pair is often correlated due to Coulomb interaction and it makes a quasi-particle known as exciton \cite{Frenkel}. Typical binding energies are a few meV in semiconductors \cite{KittelBook} but it can be as large as a few 100 meV in single-wall carbon nanotubes (SWCNTs) as the Coulomb interaction is only weakly screened in the one-dimensional surrounding \cite{AvourisPRL2004,LouiePRL2004}. Only excitons with a singlet spin wave-function (or a singlet-exciton) can be generated in the absorption of a single photon \cite{WangSCI2005,MaultzschPRB2005} but the non spin-conserving spin-orbit coupling can generate an exciton in a spin-triplet state (a triplet-exciton). This low probability process is known as intersystem-crossing (ISC) \cite{KashaRule}. The ISC results in the accumulation of triplet states, the triplet-exciton decay being a slow process as it again violates spin conservation.

Knowledge of the singlet/triplet energy structure as well as the ISC and the triplet decay dynamics is of great importance for photophysics applications. Among others the presence of triplet states hinders light harvesting due to the long recovery of the photoactive unit once it enters a triplet state. Triplet excited states also pose a challenge for light generation: an injected electron-hole pair forms a triplet state with 75 \% probability thus the theoretical limit on the internal quantum efficiency for a fluorescence based light emitting device is 25 \% (Ref. \cite{TripletOLED}). On the other hand, the ability to address and manipulate a long-living triplet state can be exploited for quantum information storage such as in the well-known NV center in diamond \cite{WrachtrupReview,AwschalomReview}. An additional motivation arises from the possible use of excitons in switching electronics \cite{KisNatPhot2019,KisNature2018}.

The photophysics of SWCNTs have been intensively studied due to their potential for light related applications and the basic optical properties are well known \cite{OConnellSCI,Bachilo:Science298:2361:(2002),WangSCI2005}. Light emission in SWCNTs arises from a bright, singlet exciton level, or $\text{S}_1$, which is one of four states due to the two-fold, K-K' degeneracy of the conduction and valence bands \cite{AvourisPRL2004,LouiePRL2004,MortimerNicholasPRB2007}. The other three exciton levels are referred to as dark since light emission from these levels is not allowed in first order by momentum and angular momentum conservation \cite{AvourisPRL2004,LouiePRL2004,WangSCI2005,MaultzschPRB2005}. The lowest lying exciton level is dark, separated by a few meV from the bright exciton level \cite{MortimerNicholasPRB2007}. This causes that the non-radiative recombination processes dominate and the radiative fluorescent process has a low probability with a quantum yield below 1 \% (Refs. \cite{LebedkinPRB,HertelJACS}). The $\text{S}_1$ bright exciton state is usually populated by irradiating the second bright singlet exciton level, $\text{S}_2$, which rapidly relaxes to $\text{S}_1$ by internal conversion. Given that both the absorption and emission energies depend strongly on the SWCNT $(n,m)$ chiral indices, the fluorescent studies have a great analytic value to characterize \textit{e.g.} the abundance of a particular SWCNT in a sample \cite{Bachilo:Science298:2361:(2002)}.

In addition to the 4 spin-singlet exciton states, each can form a spin-triplet state (referred to as triplet in the following), \textit{i.e.} altogether the SWCNTs have 16 exciton levels of which only one is bright, that gives rise to the usual band-gap fluorescence \cite{,WangSCI2005,MaultzschPRB2005}. {\color{black}Recent magneto-photoluminescence showed the influence of the triplet states on the photoluminescence emission \cite{KimACSNano}.}
The triplet levels lie below the singlet, due to the exchange interaction, which mimics Hund's rule for atoms.

Here, we report the direct observation of the phosphorescence signal from the triplet levels in pristine SWCNTs in an $(n,m)$ chiral index resolved manner. The key to this was the recent development of a unique ODMR spectrometer \cite{NegyediRSI} which operates with a tunable laser source, thus selecting a particular absorption energy, and by analyzing the wavelength of the emitted signal under magnetic resonance conditions. Another overcome challenge was the near-infrared emission range for most SWCNTs where detector efficiency is lower. Our experiment yields the singlet-triplet energy gap for several $(n,m)$ chiralities and the result is in good agreement with the theoretically predicted values in Ref.\cite{CapazPSSB}, which indicates that the exchange energy is strongly diameter dependent. Intensive microwave irradiation allowed to determine the spin-relaxation time for the triplet levels. We observe an absence of the so-called bundle-peaks in the ODMR spectra, which indicates a dominance of F\"{o}rster exciton energy transfer over conventional optical decay routes for neighboring nanotubes.

\section{RESULTS and DISCUSSION}

\begin{figure*}[htp]
\begin{center}
\includegraphics{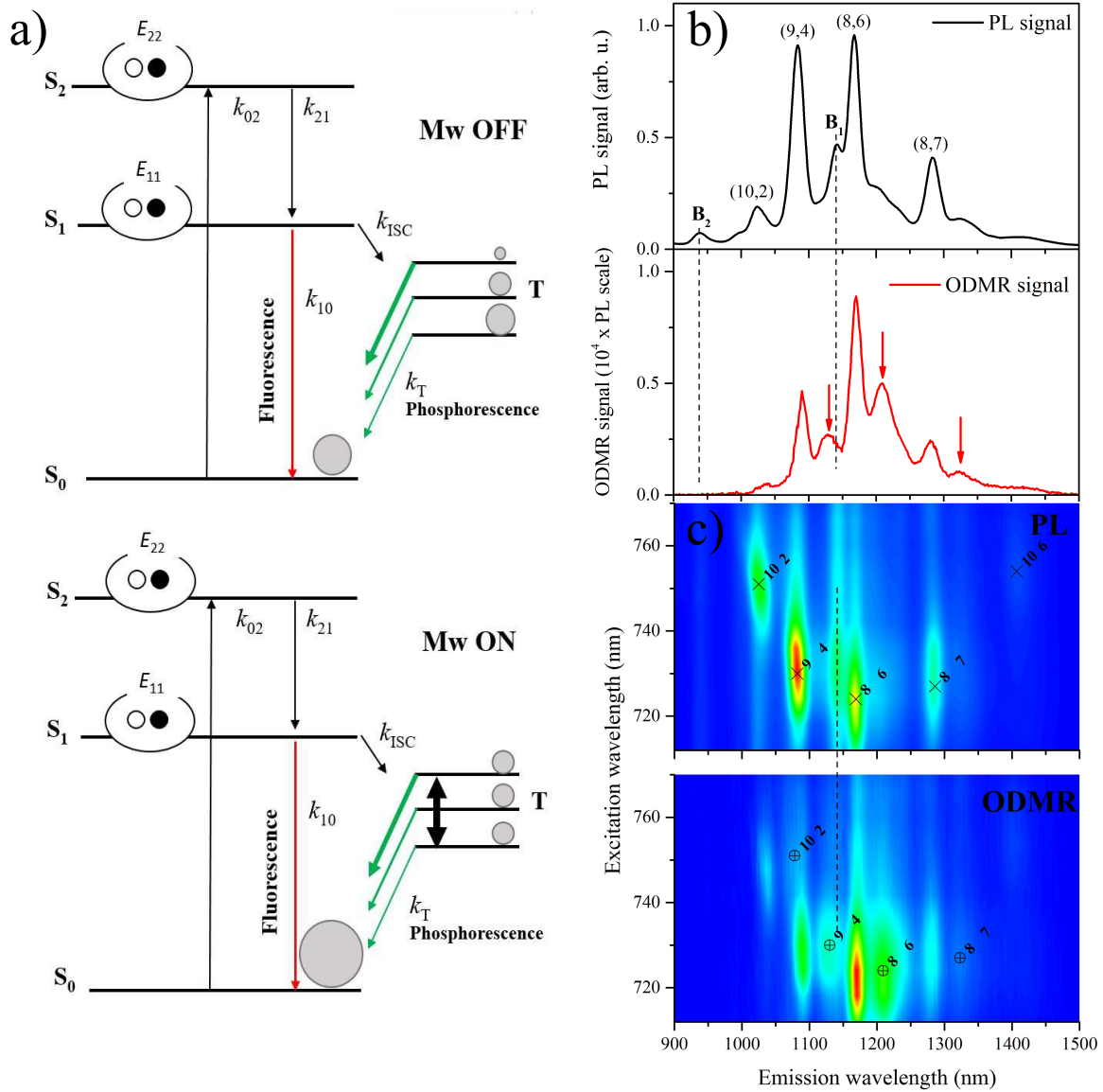}
\caption{\textbf{Jablonski diagram and ODMR results on SWCNTs.} \textbf{a}, The Jablonski diagram for the SWCNTs which illustrates the ODMR induced processes for fluorescence and phosphorescence. \textbf{b}, Individual PL and ODMR spectra when excited at 724 nm. Chiral indices are given, vertical arrows indicate the phosphorescent sidebands, vertical dashed lines highlight "bundle-peaks" (B1 and B2). \textbf{c}, PL and ODMR maps of SWCNTs. Maxima of the PL signals are indicated by $\times$ and the corresponding $(n,m)$ chiral indices are also given, $\oplus$ indicates the maxima of the phosphorescence sidebands on the ODMR map. Note the absence of the "bundle-peak" (vertical dashed lines) on the ODMR map. }
\label{fig:Fig1_Jablonski_PL_ODMR_map}
\end{center}
\end{figure*}


Fig. \ref{fig:Fig1_Jablonski_PL_ODMR_map}a. shows the Jablonski diagram of the SWCNT photophysics for the lowest lying bright exciton \cite{odmrreview1,odmrreview2,odmrreview3,odmrreview4}. Note that excited exciton states with $L\neq 0$ form a Rydberg series and these cannot be excited in single photon experiments \cite{HeinzPRL2004}. The figure illustrates that the $\text{S}_2$ bright singlet exciton level is excited from the ground state by absorption, which corresponds to an optical transition between the 2nd van Hove singularity pair \cite{OConnellSCI,Bachilo:Science298:2361:(2002)} in the single-particle description. This excited state relaxes to the $\text{S}_1$ bright exciton state by internal conversion. Its recombination to the ground state yields the fluorescence but it can also cross-over to the triplet state, $\text{T}$, due to spin-orbit interaction. The recombination of the triplet is not allowed in first order by spin-conservation, but again spin-orbit coupling can give rise to a faint phosphorescent signal. The detection of the triplet emission using magnetic resonance is called optically detected magnetic resonance (ODMR, Refs.\cite{odmrreview1,odmrreview2,odmrreview3}), which is also used herein.

Understanding the triplet energy structure, the singlet-triplet energy gap, $E_{\text{ST}}$ is important for future optoelectronics applications of SWCNT \cite{SWCNT_FET_SCI,SWCNT_Diode_SCI} or to assess the utility of the triplet level \textit{e.g.} quantum information storage potential \cite{DiamondQubitSCI,WrachtrupNatPhys2010}. In particular, it would be important to gain this information in an $(n,m)$ specific manner as a typical nanotube sample contains a large number of various SWCNTs. Spectroscopic evidence for the existence of triplet levels came from chemically modified SWCNTs with an enhanced phosphorescent activity \cite{NagatsuPRL}, more recently from ODMR studies,\cite{HertelNatPhot}{\color{black} and also from magneto-photoluminescence \cite{KimACSNano}.}

 The $E_{\text{ST}}$ for the chemically modified SWCNTs was found to be about twice as large as the theoretically predicted value\cite{CapazPSSB}. The ODMR study did not resolve the separate fluorescence and phosphorescence,\textit{i.e.} the direct observation of light emission from the triplet state in pristine SWCNTs remains elusive.



In Fig. \ref{fig:Fig1_Jablonski_PL_ODMR_map}b., we show individual PL and ODMR spectra at 77 K, when excited at 724 nm. The ODMR spectrum is the change in the optical signal due to the microwave irradiation, $\Delta\text{PL}$. Note that the PL spectrum is 4 orders of magnitude larger than the ODMR signal. The ODMR signal emerges as the $\text{S}_1$ singlet exciton state undergoes intersystem crossing to the triplet, $\text{T}$, state due to spin-orbit coupling. Recombination of the three triplet levels to the ground state can give rise to a faint phosphorescent signal due to the low probability as a result of spin-conservation. Again, spin-orbit coupling enables recombination of the triplet states. The weakness of this process hinders its direct observation in optical studies and these states are often referred to as dark triplet excitons. It was shown in Ref. \cite{NagatsuPRL} that hydrogen coverage of SWCNTs can enhance a sideband signal which was associated with the phosphorescence of SWCNTs. The master equation for the triplet population, $\left[\text{T} \right]$ reads:

\begin{gather}
\frac{\text{d}\left[\text{T}\right]}{\text{d}t}=k_{\text{ISC}}\left[\text{S}_1\right]-k_{\text{T}}\left[\text{T}\right],
\label{eq:triplet_master}
\end{gather}
where $\left[\text{S}_1\right]$ is the population of the excited state. Under steady state conditions, Eq. \eqref{eq:triplet_master} leads to a finite triplet population, with the photoactive states being effectively "trapped" in the long-living triplet state. This means that ISC not only gives rise to a finite triplet population but it also reduces the fluorescent signal. 

{\color{black}In principle, the ISC process can be reversible due to triplet-triplet interactions at high irradiation fluences \cite{HertelNatPhot} or parity flipping due to adsorbed molecules \cite{IshiiPRX2019}. In our case, neither of these mechanisms can be significant and our wavelength resolved approach enables to discriminate to origin of the emitted photon, as shown below.}

Fig. \ref{fig:Fig1_Jablonski_PL_ODMR_map}a. illustrates the origin of the ODMR signal as follows. The three sublevels of the triplet are split either by zero-field splitting (ZFS) or by an external magnetic field (the Zeeman effect). The presence of a strong ZFS is typical for strongly localized (Frenkel) excitons \cite{odmrreview2}, however when ZFS is small, which is the case for a larger Mott-Wannier type exciton such as in SWCNTs\cite{DoornJACS}, an external magnetic field can split the levels. The population of the three sublevels is usually unequal immediately after ISC as they couple differently to the lattice due to their differing orbital arrangement. Similarly, the decay rates to the ground states of the individual triplet sublevels are also usually different. An intensive microwave irradiation on magnetic resonance condition equilibrates the populations of the triplet sublevels, thereby affecting the phosphorescent signal. A consideration of the number of available states \cite{odmrreview2} yields (also discussed in the Supporting Information) that the microwave induced additional phosphorescent flux \emph{equals} the additional fluorescent flux, \textit{i.e.} the more states are liberated from the triplet state, the more fluorescent transitions can occur. We note that our signal is positive, \textit{i.e.} the presence of microwave irradiation increases both the fluorescent and the phosphorescent signals.
A previous ODMR report in SWCNTs did not resolve these two processes as no wavelength analysis of the emitted light was performed \cite{HertelNatPhot}. 

For SWCNTs, both the $\text{S}_0 \rightarrow \text{S}_2$ absorption and the $\text{S}_1 \rightarrow \text{S}_0$ fluorescence emission depend on the $(n,m)$ indices \cite{OConnellSCI}, which allowed a chiral index assignment of the individual fluorescence peaks \cite{Bachilo:Science298:2361:(2002)}. When plotted as a function of excitation and emission, the individual PL spectra yields a PL map, which is shown in Fig. \ref{fig:Fig1_Jablonski_PL_ODMR_map}c. The individual ODMR spectra can be combined in a similar manner, which yields the ODMR map in Fig. \ref{fig:Fig1_Jablonski_PL_ODMR_map}c. Resolution is 1 nm for the emission and 6 nm for the excitation. The PL map \cite{OConnellSCI} contains well-known features, including some strong emission peaks at the corresponding excitation energy, which are indicated with "$\times$" in the figure and the corresponding $(n,m)$ chiral indices are also given. 

The emission and excitation energies in our PL map slightly differ from well-known literature values \cite{WeismanNL2003}, due to a differing surfactant, temperature, or the level of bundling. However, SWCNT chiralities could be unambiguously identified due to the patterned nature of the transition energies\cite{OConnellSCI,Bachilo:Science298:2361:(2002)}. In addition, the PL map also contains features which are colloquially known as "bundle-peaks" \cite{bundle1,bundle2,DoornJACS} and are indicated by vertical solid lines. These features appear when a few nanotubes form a small and interacting aggregate structure and an SWCNT with appropriate transition energy absorbs light. The resulting exciton is then transferred to another SWCNT in the bundle, which then emits at its characteristic emission wavelength \cite{bundle1,bundle2}. This process leads to a characteristic feature of the bundle peaks that they appear as vertical lines in the PL maps. These peaks are \emph{absent} in the ODMR map, which is discussed further below.

\begin{figure}[htp]
\includegraphics[width=0.48\textwidth]{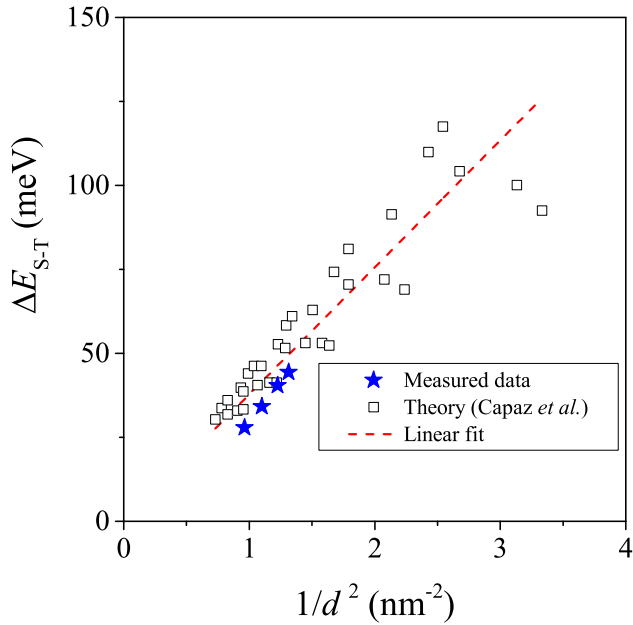}
\caption{\textbf{Singlet-triplet gap in SWCNTs.} Magnitude of the singlet-triplet energy gap for the four SWCNTs which were identified in our study ($\star$) as a function of the inverse of the squared nanotube diameter, $1/d^2$. The theoretical prediction by Capaz \textit{et al.} (Ref. \cite{CapazPSSB}) is shown with $\Box$ symbols along with a linear fit to the data.}
\label{fig:Fig2_Delta_ST}
\end{figure}

We note that the main fluorescent peaks do not appear at exactly the same position on the PL and ODMR studies beyond the experiment accuracy. We believe that this effect is due to the known inhomogeneous broadening of the PL lines \cite{OConnellPRL2004} either due to a different charging, defects, or a slightly different environment. We speculate that the ISC transition rates and thus the triplet population is affected by these factors, \textit{i.e.} the microwave induced fluorescent peaks in the conventional PL studies should not necessarily match those in the ODMR results. However, this effect requires further study. 

 An exception is the (10,6) peak which should show a satellite in the wavelength region where our spectrometer sensitivity abruptly drops. It is interesting to note that the phosphorescent peaks have the same integrated intensity as the fluorescent ones but are broader. The same intensity is explained above and the larger linewidth could be related to a larger sensitivity of the triplet level to the above mentioned environmental factors. We note that the Zeemann splitting of the triplet levels (9 GHz) is much smaller than the extra broadening (1 nm at 1100 nm wavelength corresponds to 250 GHz). 


The observation of the phosphorescence in ODMR for four SWCNT chiralities allows to determine the \emph{chirality resolved} singlet-triplet energy gap, $\Delta_{\text{ST}}$. The result is shown in Fig. \ref{fig:Fig2_Delta_ST}. together with the theoretical prediction by Capaz \textit{et al.} (Ref. \cite{CapazPSSB}). The latter calculation was based on considering the variation of the exchange interaction due to the finite size quantum confinement. 

{\color{black}The exciton size scales with the tube diameter, $d$, as it is the only length scale in the problem. The exciton binding energy scales with $1/d$ as the Coulomb interaction energy scales with the inverse distance. 
The singlet-triplet splitting is due to the exchange energy and the latter varies as $1/d^2$ as it is due to the electric dipole self-interaction of a neutral charge distribution\cite{CapazPRB2006}.} {\color{black}A similar, $1/d^2$ dependence was observed for the energy difference between the bright and defect-brightened, otherwise dark excitons\cite{PiaoNatChem}. This is due to the fact that the bright-dark exciton splitting is also related to the exchange interaction according to Ref. \cite{CapazPRB2006}.}

Our experimental data matches well with the theoretical prediction which indicates that the theoretical model was indeed appropriate. This agreement also {\color{black}enables the future} identification of phosphorescence in SWCNTs with chiralities which are not presented herein. {\color{black}Certainly, extending our measurements toward SWCNTs with largely different diameters (such as \textit{e.g.} the (6,4) and (6,5) SWCNTs) would improve the confidence in the agreement.}

\begin{figure}[htp]
\includegraphics[width=0.45\textwidth]{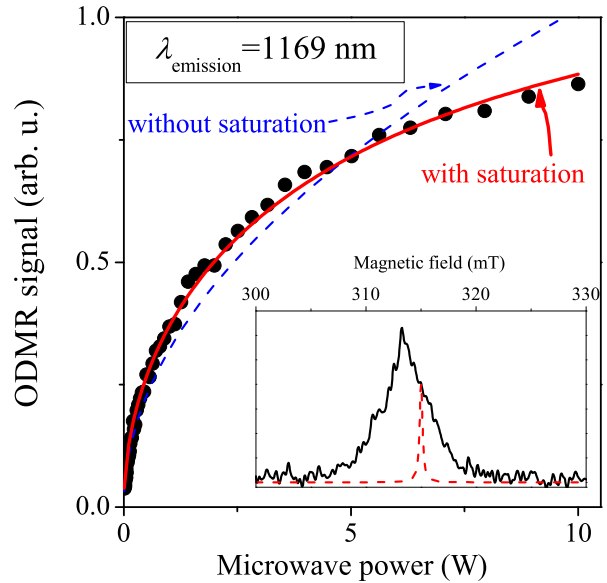}
\caption{\textbf{Microwave power dependence of the ODMR signal.} The emission wavelength is 1169 nm that corresponds to the (8,6) SWCNT. Dashed and solid curves show a fit without and with saturation effects included, respectively. The inset shows a magnetic field swept ODMR spectrum and the signal of an individual spin-packet is also indicated with a dashed line.}
\label{fig:Fig3_saturation}
\end{figure}

We studied the microwave power, $p$, dependence of a selected peak (of the (8,6) SWCNT) at 1169 nm emission and 724 nm excitation in the ODMR map, in order to gain information about the characteristic spin-dynamics timescales: $T_1$ (spin-lattice relaxation time) and $T_2$ (spin-spin relaxation time). The magnitude of these parameters determines whether the SWCNT triplet states could be used \textit{e.g.} for quantum information storage purposes.

The power dependence of the ODMR signal is shown in Fig. \ref{fig:Fig3_saturation}. The ODMR signal is expected to follow the population change in the triplet sublevels, induced by the microwave excitation. For low microwave powers, it follows the strength of the microwave magnetic field\cite{SlichterBook}, $B_1$: $\text{ODMR}\,\propto{B_1}\propto\sqrt{p}$, however at sufficiently high powers, a saturation could be observed as\cite{SlichterBook}: 

\begin{equation}
\text{ODMR}\,\propto \sqrt{\frac{p}{1+C p Q  \gamma^2 T_1 T_2}}.
\label{Eq:Saturation}
\end{equation}
Herein, $Q=1,000$ is the quality factor of our microwave cavity, $C=2.2\cdot 10^{-12}\,\text{T}^2/\text{W}^2$ is a resonator constant which links the generated microwave magnetic field strength to the power and is resonator mode dependent\cite{PooleBook} (it is the cylindrical TE011 mode in our case). Besides the gyromagnetic ratio, $\gamma/2\pi=28\,\text{GHz/T}$, the formula allows to obtain the $T_1 T_2$ product\cite{PortisPR1953} when saturation is present. 

Fig. \ref{fig:Fig3_saturation}. shows that a fit to the power dependent ODMR signal without saturation does not account for the data at high microwave powers. When saturation is included, Eq. \eqref{Eq:Saturation} describes well the data in the whole power range and we obtained $\sqrt{T_1 T_2}=36(1)\,\text{ns}$. Although, the data itself does not allow to disentangle the values of $T_1$ and $T_2$ separately, we argue below that most probably $T_2=T_1$ holds due to the short value of $T_1$. This case can be related to a homogeneous linewidth of $\Delta B_{\text{hom}}=1/\gamma T_1=0.16\,\text{mT}$.

In general the $T_2 < T_1$ hierarchy holds, except when $T_1$ becomes so short that it shortens the $T_2$ as well\cite{FabianRMP} which results in $T_2=T_1$. For most materials, $T_2$, and the related linewidth $\Delta B_{\text{dipole}}=1/\gamma T_2$, is due to the dipole-dipole interaction between like spins, which can be quantified using the van Vleck formula \cite{AbragamBook}. {\color{black}A significant exciton-phonon interaction \cite{YoshikawaAPL} could also lead to additional dephasing and thus a shorter value of $T_2$. However, it is expected that at low temperatures, this effect vanishes and the dipole-dipole interaction dominates.} A reasonable estimate for the maximum value of $\Delta B_{\text{dipole}}=0.02\,\text{mT}$ can be obtained for a triplet exciton to triplet exciton distance of 10 nm, which is close to the exciton Bohr radius, thus it can be regarded as a lower limit for their separation \cite{MaultzschPRB2005}. However, this value of $\Delta B_{\text{dipole}}$ is smaller than the above $\Delta B_{\text{hom}}$ by an order of magnitude. This means that dipole broadening cannot account for the observed $T_1 T_2$ value, we are thus led to conclude that $T_1=T_2=36(1)\,\text{ns}$ at 77 K in the SWCNTs. 

We note that the spectral linewidth of a magnetic field swept ODMR spectrum is about 5 mT (see the inset of Fig. \ref{fig:Fig3_saturation}), which is much larger than the above mentioned homogeneous linewidth. Portis showed\cite{PortisPR1953} that it is the linewidth of the so-called spin-packets with linewidth $1/\gamma T_2$ (shown with a dashed curve in the inset of Fig. \ref{fig:Fig3_saturation}.) that enters  the saturation formula and that the visible linewidth (or signal envelope) does not affect it. As per the above short values of $T_1$ and $T_2$, we believe that it is caused by the electron-phonon interaction, which is mediated to the spin state through spin-orbit coupling \cite{FabianRMP} but its origin requires additional theoretical studies. This short value of $T_{1,2}$ appears to be prohibitively short if the triplet state would be used to store quantum information. Although SWCNTs appear particularly advantageous for such purposes \cite{WrachtrupNatPhys2010}, given the compatibility of the emission and absorption wavelengths with the telecommunication windows ($\approx 1300-1550\,\text{nm}$), a $T_1$ of a few ms or $\mu$s is required for a meaningful operation\cite{BudkerPRL2012,WrachtrupNatPhys2010}.

\begin{figure*}[htp]
\begin{center}
\includegraphics[width=0.95\textwidth]{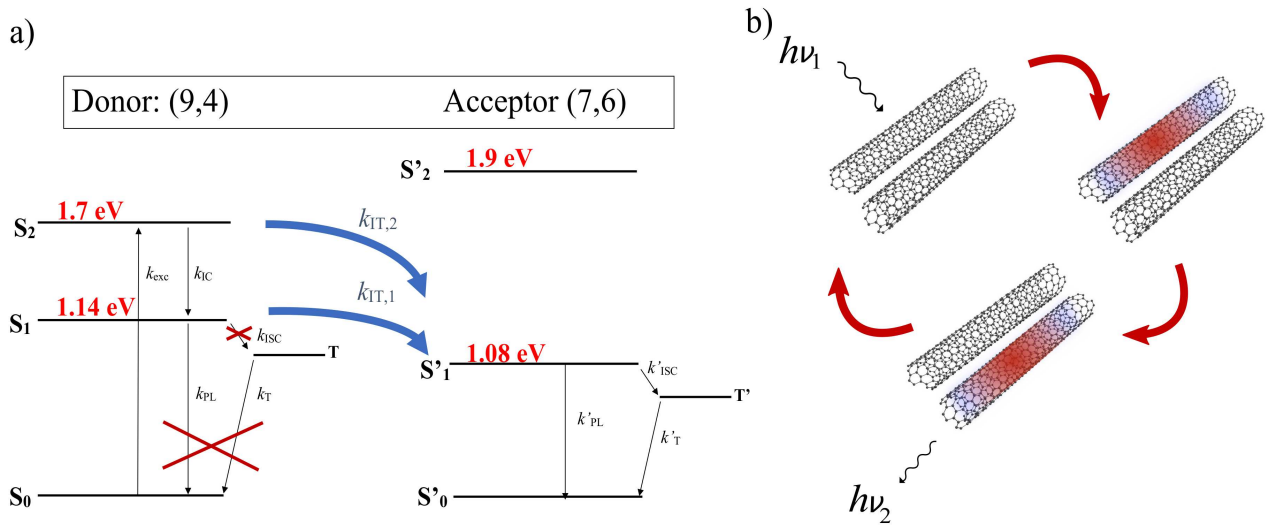}
\caption{\textbf{Jablonski diagram (a) and schematics (b) for two neighboring SWCNTs.} The exciton created on the donor SWCNT (the (9,4) in this case) is transferred to the neighboring acceptor tube (the (7,6) tube in this case) with a high probability (thick arrows) thus the usual relaxation processes (crossed-out arrows) do not occur. Schematic depiction of the dominating exciton energy transfer process: the absorbed light forms an exciton which is then transferred to the acceptor SWCNT where the exciton recombination and light emission occurs.}
\label{fig:Fig4_bundle}
\end{center}
\end{figure*}

We mentioned above that the "bundle-peaks", B$_1$ and B$_2$, are absent in the ODMR spectra as Fig. \ref{fig:Fig1_Jablonski_PL_ODMR_map}c. demonstrates. As mentioned, these are the emission peaks which are related to a F\"{o}rster exciton energy transfer process between two neighboring nanotubes which form a small bundle. This absence in the ODMR cannot be due to experimental factors: the B$_2$ peak falls well within the range of sensitivity of our system and this spectral range is free of other SWCNT peaks, still the ODMR spectrum shows no trace of a peak. Concerning the B$_1$ peak, the spectral range has more peaks, however, the PL map shows that this bundle-peak is particularly strong and has a comparable intensity to the (9,4) and (8,6) peaks. However, the ODMR spectrum contains no signature of these peaks. 

The relevant Jablonski diagram for the B$_1$ line in Fig. \ref{fig:Fig1_Jablonski_PL_ODMR_map}c. is shown in Fig. \ref{fig:Fig4_bundle}. Here, the donor SWCNT is the (9,4) and the exciton is transferred to the acceptor (7,6) which then radiates the observed emission (thick curved arrows in the figure). It cannot be decided from the present data whether the exciton energy transfer occurs from the S$_2$ (intertube transfer process, $k_{\text{IT,1}}$) or S$_1$ (intertube transfer process: $k_{\text{IT,2}}$) levels of the (9,4) donor.  

Our interpretation for the absence of the bundle-peaks in the ODMR data is that for neighboring nanotubes the exciton transfer process dominates over the usual relaxation processes. In other words, whenever possible the exciton energy transfer occurs. In case of the B$_1$ line, the exciton is transferred with a probability of 1 from the (9,4) to the (7,6) SWCNT, which does not allow the population of the triplet state, T, on the (9,4) nanotube. This explains the insensitivity of the bundle-peak to the ODMR conditions. 

The dominance of the F\"{o}rster exciton energy transfer process was predicted in Ref. \cite{LefebvreJPhysChem} due to the sizeable interaction between SWCNTs in a small bundle and the overlapping of optically active states but to our knowledge the magnitude of its probability has not been quantified yet. The F\"{o}rster exciton energy transfer process is known to play an important role in so-called molecular ruler applications \cite{ForsterMolecularRuler} due to the sensitivity of the process for molecular distances and also for fundamental exciton transfer processes including photosynthesis \cite{ForsterBook}. Our work thus {\color{black}allows} to explore the role of SWCNTs during applications which involve the exciton energy transfer either from or to SWCNTs.

\section{CONCLUSIONS}
In conclusion, we studied the optically detected magnetic resonance spectra of SWCNTs by selectively exciting the absorption bands and by analyzing the emitted signal. The study allowed the direct observation of phosphorescence from chirality assigned SWCNTs. The SWCNT diameter dependent singlet-triplet energy gaps were obtained and it was found to agree well with the value of theoretical predictions and the expected $1/d^2$ diameter dependence was found. A saturation experiment with varying microwave power allowed to determine the nominal spin-lattice relaxation time of the triplet sublevels. The bundle-peaks are absent in the ODMR spectra which is explained by the dominance of the F\"{o}rster exciton energy transfer between neighboring nanotubes over the usual relaxation processes of excitons. We believe that these findings {\color{black}may enable} quantum information storage, manipulation and readout using telecom compatible wavelengths while using carbon nanotubes.

\section{METHODS}

Single-wall carbon nanotubes were studied in aqueous suspensions following the conventional routes \cite{OConnellSCI,Bachilo:Science298:2361:(2002)}. A solution of sodium-deoxycholate (DOC) surfactant and distilled water ($2\,\%$ concentration) was mixed with HiPco nanotubes (Carbon Nanotechnologies Inc.) in an ultrasonic bath. This type of nanotube is known to contain several SWCNT chiralities with a mean diameter of 1 nm (Ref. \cite{KukoveczEPJB}). 

The nanotube and surfactant mixture was intensively sonicated with a tip-sonicator (Branson Sonifier 450) for one hour with $60\%$ output intensity. During the sonication, the suspension is cooled with a circulating water system. This process makes the nanotubes individualized as the surfactant molecules envelop the nanotubes and help their de-bundling. The next step is an ultracentrifugation of the samples with $200\,\text{kg}$ acceleration for an hour. During this part of the preparation, the nanotubes which are not individualized, settle down at the bottom of the cuvette. Finally, the sample is decanted from the top of the suspension and placed into a 4 mm diameter quartz tube for the measurements.

The development of the ODMR instrument, which allows energy resolved excitation and detection in the near infrared domain, was reported in Ref. \cite{NegyediRSI}. A tunable laser system, comprising of a 532 nm laser (5 W) pumped Ti:Sa laser enables tunable excitation of the SWCNTs and emitted light is detected in a dispersive monochromator system (Horiba Jobin Yvon ihr320) using a liquid nitrogen cooled InGaAs detector in the 900-1500 nm range. The samples are in a liquid nitrogen cryostat inside a TE011 microwave cavity which is driven by a TWT amplified microwave signal, whose intensity is chopped. The microwave induced optical signal is measured in phase with the chopped microwaves using a lock-in amplifier, which yields the ODMR signal or $\Delta\text{PL}$. The DC component of the optical signal yields the conventional PL signal and the two types of signals are measured simultaneously.

\section{AUTHOR INFORMATION}
\textbf{Author Contributions}\\
JP and MN contributed equally to this work. MN and FS developed, JP, SK, and AB tested extensively the ODMR spectrometer. MN performed the ODMR studies and JP analyzed the data. PR and TP prepared and characterized the SWCNT samples. FS and JP developed the kinetic models. All authors contributed to writing the manuscript.

\smallskip

\noindent\textbf{Notes}\\
The authors declare no competing financial interest.

\section{ACKNOWLEDGEMENTS}
Work supported by the Hungarian National Research, Development and Innovation Office (NKFIH) Grant Nrs. 2017-1.2.1-NKP-2017-00001 and K119442, the FP7-GA-607491 (COMIQ), {\color{black}and by the FWF (P27769-N20) grants. FS acknowledges the hospitality of L. Forr\'{o} during a sabbatical stay and financial support by the Swiss National Science Foundation (Grant 200021 144419) and European Research Council (ERC) Advanced Grant Nr. 670918.}


\providecommand{\latin}[1]{#1}
\makeatletter
\providecommand{\doi}
{\begingroup\let\do\@makeother\dospecials
	\catcode`\{=1 \catcode`\}=2 \doi@aux}
\providecommand{\doi@aux}[1]{\endgroup\texttt{#1}}
\makeatother
\providecommand*\mcitethebibliography{\thebibliography}
\csname @ifundefined\endcsname{endmcitethebibliography}
{\let\endmcitethebibliography\endthebibliography}{}

\end{document}